\documentclass[sigconf,nonacm]{acmart}

\usepackage{colortbl}
\usepackage{orcidlink}
\AtBeginDocument{%
  }

\setcopyright{acmlicensed}
\copyrightyear{2025}
\acmYear{2025}
\acmDOI{XXXXXXX.XXXXXXX}
\acmISBN{978-1-4503-XXXX-X/2018/06}

\acmConference[EASE 2025]{The 29th International Conference on Evaluation and Assessment in Software Engineering}{17--20 June, 2025}{Istanbul, Turkey}

\begin{document}

\title{A Study on Mixup-Inspired Augmentation Methods
for Software Vulnerability Detection}


\author{Seyed Shayan Daneshvar~\orcidlink{0000-0002-3463-7873}}
\orcid{0000-0002-3463-7873}
\affiliation{%
  \department{Department of Computer Science}
  \institution{University of Manitoba}
  \city{Winnipeg}
  \state{MB}
  \country{Canada}
}
\email{daneshvs@myumanitoba.ca}

\author{Da Tan~\orcidlink{0009-0000-7861-6786}}
\orcid{0009-0000-7861-6786}
\affiliation{%
  \department{Department of Computer Science}
  \institution{University of Manitoba}
  \city{Winnipeg}
  \state{MB}
  \country{Canada}
}
\email{tand2@myumanitoba.ca}

\author{Shaowei Wang~\orcidlink{0000-0003-3823-1771}}
\orcid{0000-0003-3823-1771}
\affiliation{%
  \department{Department of Computer Science}
  \institution{University of Manitoba}
  \city{Winnipeg}
  \state{MB}
  \country{Canada}
}
\email{shaowei.wang@umanitoba.ca}

\author{Carson K. Leung~\orcidlink{0000-0002-7541-9127}}
\orcid{0000-0002-7541-9127}
\affiliation{%
  \department{Department of Computer Science}
  \institution{University of Manitoba}
  \city{Winnipeg}
  \state{MB}
  \country{Canada}
}
\email{carson.leung@umanitoba.ca}

\renewcommand{\shortauthors}{Daneshvar et al.}

\begin{abstract}
Various deep learning (DL) methods have recently been utilized to detect software vulnerabilities. Real-world software vulnerability datasets are rare and hard to acquire, as there is no simple metric for classifying vulnerability. Such datasets are heavily imbalanced, and none of the current datasets are considered huge for DL models. To tackle these problems, a recent work has tried to augment the dataset using the source code and generate realistic single-statement vulnerabilities, which is not quite practical and requires manual checking of the generated vulnerabilities. In this paper, we aim to explore the augmentation of vulnerabilities at the representation level to help current models learn better, which has never been done before to the best of our knowledge. We implement and evaluate five augmentation techniques that augment the embedding of the data and have recently been used for code search, which is a completely different software engineering task. We also introduced a conditioned version of those augmentation methods, which ensures the augmentation does not change the vulnerable section of the vector representation. We show that such augmentation methods can be helpful and increase the F1-score by up to 9.67\%, yet they cannot beat Random Oversampling when balancing datasets, which increases the F1-score by 10.82\%. 
\end{abstract}




\maketitle

\section{Introduction}
In software engineering and development, software vulnerabilities are widespread, which are the cause of many security risks in the industry. To help with this problem, researchers have tried to come up with models that can predict and classify the vulnerability of code pieces \cite{text_mining_vul, LineVD, IVDetect, DeepVD, fu_linevul_2022,vuldeepecker, Devign, Reveal}. However, all of these works face the reality of being unable to perform well in real-life scenarios due to a shortage of data. Chakraborty et al.~\cite{Reveal} revealed a substantial 73\% average drop in F1-scores for state-of-the-art (SOTA) deep learning-based vulnerability detection (DLVD) models when dealing with real-world imbalanced data. This performance decline poses a challenge to the practical application of DLVD, prompting the exploration of strategies like data sampling~\cite{yang_does_2023}, data augmentation and generation~\cite{nong_vulgen_2023, vgx, Vulscriber}, cost-sensitive learning~\cite{cost-sensitive}, and ensemble methods~\cite{boosting}. 

To deal with the shortage of vulnerable samples, Nong et al.~\cite{nong_vulgen_2023, vgx} proposed limited solutions (i.e., VGX and Vulgen) for generating single statement vulnerabilities using a deep learning model and a data mining-based method. Yang et al.~\cite{yang_does_2023} explored the effect of different sampling methods both at the raw level and latent level, and found that the rudimentary Random Oversampling (ROS) achieves the best performance and more complex methods, such as SMOTE ~\cite{SMOTE} cannot keep up with ROS. Recently, Daneshvar et al.~\cite{Vulscriber} showed that all the previous methods have limited effectiveness and similar to SMOTE, Vulgen and VGX cannot beat ROS. They also proposed a vulnerable code generation pipeline, heavily based on prompting LLMs, that could beat ROS but was costlier than the previous methods, and does not work without using powerful LLMs.

Representation-level augmentation methods' effectiveness has been shown for various software engineering tasks---namely, code search \cite{li_exploring_2022}, problem classification \cite{MixCode, dong_boosting_2023}, bug detection \cite{MixCode, dong_boosting_2023}, authorship attribution \cite{dong_boosting_2023}, and clone detection \cite{dong_boosting_2023}. However, the efficacy of such methods remains unexplored for the task of vulnerability detection. While this task is not directly related to bug detection, it is closer than others, yet significantly more challenging. Vulnerable code functioning correctly adds complexity, requiring increased expertise and focus in real-life scenarios. Additionally, vulnerability detection datasets are smaller and more sparse. It is important to note that the nature of vulnerability datasets and the complexity of the task itself makes it imprudent to assume the effectiveness of these methods for vulnerability detection without conducting a systematic evaluation.

Hence, following the recent work on augmenting data for more common code tasks by Dong et al. \cite{MixCode} and on code search by Li et al. \cite{li_exploring_2022}, we explore and test if the text-based embedding augmentation techniques. These techniques are all influenced by the idea of MixUp \cite{Senmixup}, help with the training of token-based vulnerability detection models---such as LineVul \cite{fu_linevul_2022}---and whether they can beat ROS. We also use VGX as another baseline and show that these methods are more suitable than VGX for large-scale data augmentation.

In this paper, we explore the effect of five representation-level augmentation methods (Linear Interpolation, Stochastic Perturbation, Linear Extrapolation, Binary Interpolation, and Gaussian Scaling). Uniquely adapting these techniques for vulnerability detection, we exclusively augment the vulnerable items. Additionally, we explore a variant of these methods by conditioning the augmentation, keeping the vulnerable part of the code fixed. This approach is non-intrusive as it does not necessitate extensive source code analysis. We evaluate the data augmentation methods on LineVul~\cite{fu_linevul_2022} (a SOTA token-based DLVD) with BigVul~\cite{fan_cc_2020} dataset, and compare the results. The results show that the conditioned variation performs better in general, as it provides a more targeted enhancement to vulnerability detection by ensuring the generated embeddings contain vulnerable lines. However, ROS generates a better performance. More specifically, ROS boosts the F1-score performance by 10.82\% while the best-performing augmentation method (i.e., naive Stochastic Perturbation) boosts the performance by 9.96\%.
%
%
%
Hence, {\em key contributions} of our paper include:
\begin{itemize}
    \item To our knowledge, we conducted the first systematic and thorough study to understand how representation-level augmentation affects DLVD. We examined the impact of five data augmentation methods on a leading token-based DLVD approach, utilizing one of the largest datasets available that provides line-level information. Line-level information is crucial, particularly for the conditioned approach.
    \item We present a novel application of these augmentation methods designed for datasets that incorporate line-level information about the location of vulnerable statements. In this approach, we specifically locate and keep the representation of the corresponding vulnerable line(s) fixed during the augmentation process. This ensures that the augmented vector is more likely to reflect vulnerable data.
    \item We comply with the SIGSOFT Open Science Policy by putting our code in a permanent repository---namely, \href{https://zenodo.org/records/13916933}{Zenodo}---to the extent ethically and practically possible 
    for replication and further improvements. 
\end{itemize} 
The remainder of this paper is organized as follows. The next section introduces background and related works to DLVD and representation-level augmentation. Section~\ref{sec3} describes our experimental design, and Section~\ref{sec4} shows results. Sections~\ref{sec5}, \ref{sec6} and \ref{sec7} provide discussions, conclusions, and code availability, respectively.

\section{Background and Related Works}

\subsection{Overview of Deep Learning-Based Vulnerability Detection}
Generally, DLVD is done in three phases: feature extraction, model training, and model deployment. In the feature extraction phase, various features are extracted from the code that can effectively capture the semantic and syntactic properties of the code. Following this, the extracted features are embedded into a real-valued vector, representing the code snippet. Subsequently, these representations are used to train the model that predicts the vulnerability of a representation. For instance, LineVul~\cite{fu_linevul_2022} is a SOTA method that leverages CodeBERT \cite{feng_codebert_2020} for embedding code snippets (both vulnerable and clean samples). Finally, in the deployment phase, the same feature extraction and embedding techniques are used to transform code snippets of software systems and feed them to the model to predict the vulnerability.

\subsection{Token-Based Vulnerability Detection and LineVul} 
Derived from the extraction and embedding processes, two classes of DLVD can be identified: Token-based and graph-based. In our study, we focus on the former. In token-based DLVD approaches, each code snippet is transformed into a sequence of learnable vectors known as tokens. The token representations of the code are acquired through text embedding techniques such as Word2Vec~\cite{mikolov_efficient_2013} and BERT~\cite{devlin_bert_2019}. For instance,  LineVul~\cite{fu_linevul_2022} is a recent transformer-based~\cite{vaswani_attention_2023} SOTA method that predicts the vulnerability of functions and can locate the vulnerable lines.

\subsection{Representation-Level Augmentation} \label{sec:aug_methods}
Data augmentation for source code can be approached in various ways ~\cite{dong_boosting_2023, bui_transformation_2021, MixCode}. However, many of these methods are designed specifically for source code and necessitate source code analysis. This process is time-consuming and involves altering the code meticulously to ensure the preservation of code semantics, such as Variable Renaming and Switch to If \cite{bui_transformation_2021}. An alternative to augmenting source code directly is representation-level augmentation. This method involves augmenting data after the embedding process and manipulating the representation vectors using mathematical processes to generate new data points. Representation-level augmentation can be applied to both text and graph data. 


\begin{figure}[!t]
\centering
\includegraphics[width=\linewidth]{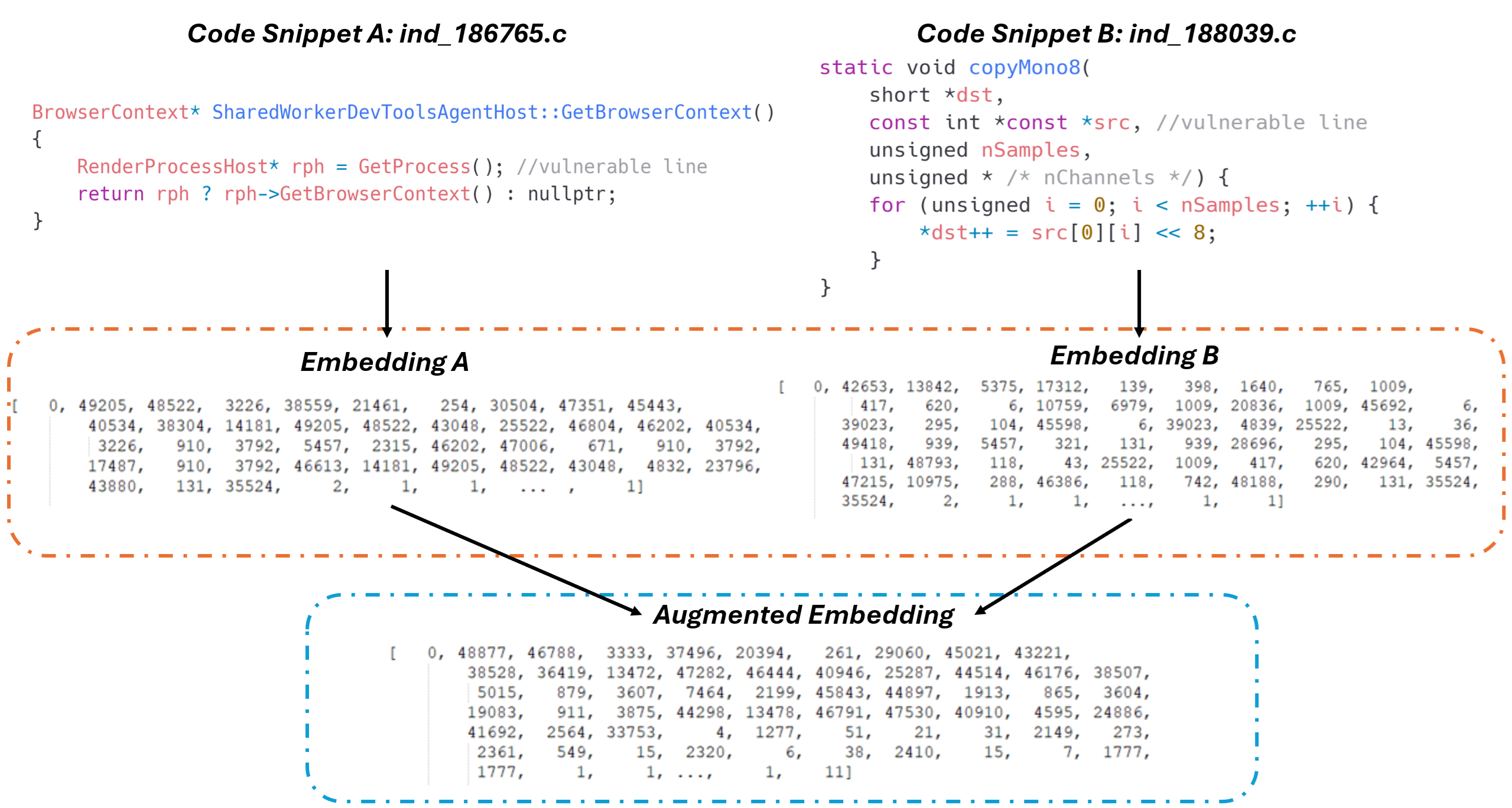}
\caption{Blind augmentation of vulnerabilities using Linear Interpolation, where the augmented embedding is a weighted average of the two embeddings.}
\label{fig:blind_aug}
\end{figure}

\begin{figure}[!t]
\centering
\includegraphics[width=\linewidth]{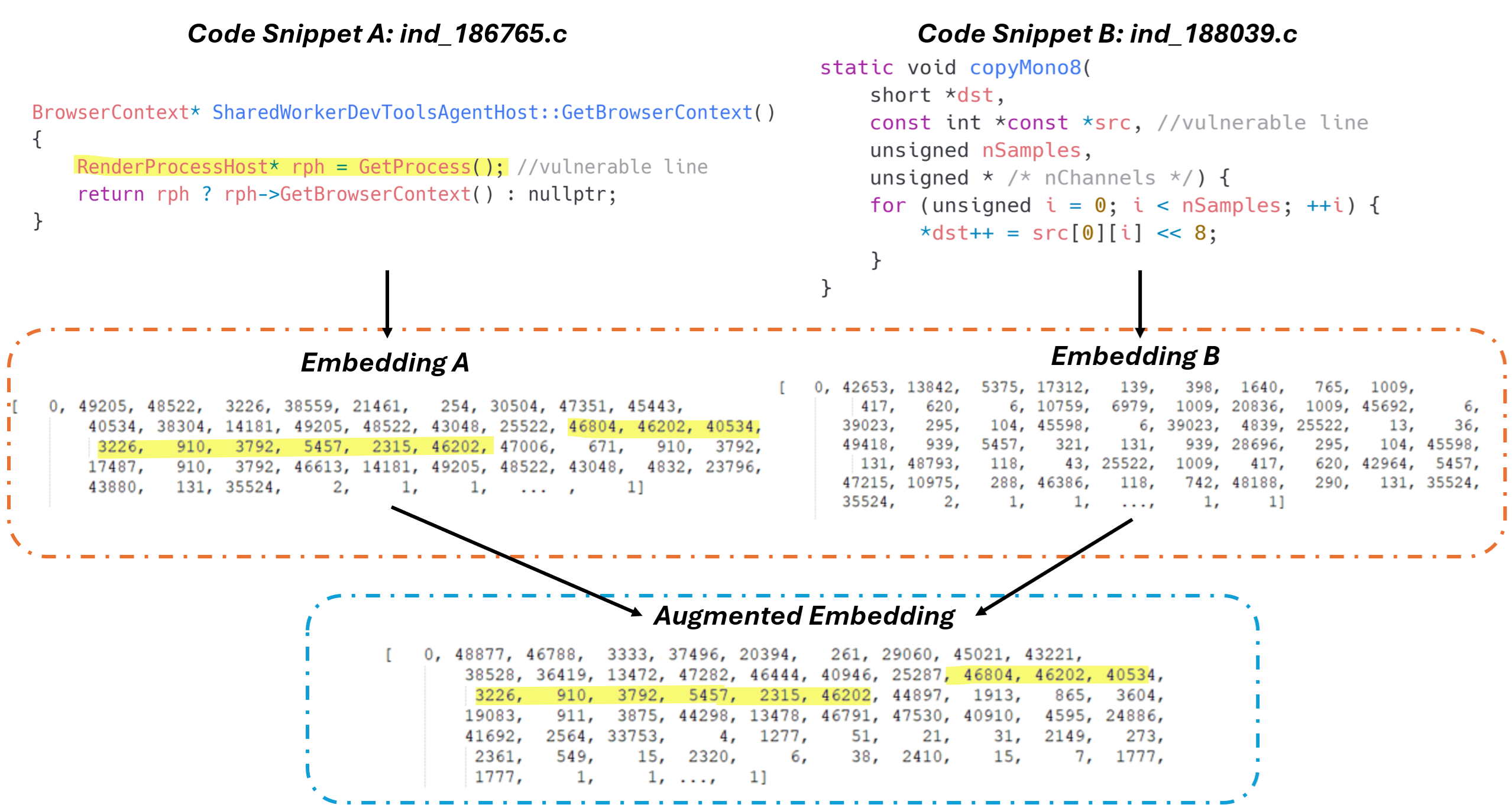}
\caption{Conditioned augmentation of vulnerabilities using Linear Interpolation. The vulnerable line and the corresponding tokens are highlighted.}
\label{fig:cond_aug}
\end{figure}

In this section, we delve into the five token-based representation-level augmentation methods following previous studies~\cite{li_exploring_2022, dong_boosting_2023, MixCode}. These methods encompass the entire spectrum of representation-level augmentation explored for source code at the text level. All five methods can be expressed using the following general format:
\begin{equation}
    h^{+} = \alpha \odot h + \beta \odot h' \label{eq:aug}
\end{equation}
where $h^{+}$ is the augmented data, $\alpha$ and $\beta$ are coefficients determined by the method, $h$ and $h'$ are two data points in their representation vector format, drawn from the original dataset. Based on the general format, these augmentation methods are described as follows:

\begin{itemize}

    \item \textbf{Linear Interpolation (LI)}, which is the initial method used for representation-level augmentation of source code and was inspired by SenMixup~\cite{Senmixup} in NLP. Hence, it may be called Mixup~\cite{Zhang2017mixupBE} for source code as well. In this method, $\alpha$ is sampled from a uniform distribution $\alpha \sim U(a, 1.0)$ for large values of $a$ and $\beta = 1 - \alpha$, and $h \neq h'$. The application of this method is depicted in Figure \ref{fig:blind_aug}, and our conditioned variation of it in Figure \ref{fig:cond_aug}.
    
    \item \textbf{Stochastic Perturbation (SP)}, which randomly deactivates features in the representation vector. It is usually implemented using DropOut, which drops features with a probability of $p$ and then multiplies the resulting vector by a factor of $\frac{1}{1-p}$. Hence, in the case of SP, $\alpha$ is sampled from a Bernoulli distribution $B(p)$, and $p$ is the probability parameter. Consequently, $\alpha \in \left\{0, \frac{1}{1-p} \right\}$, $\beta = \frac{1}{1-p} - \alpha$, and $h' = 0$.

    \item \textbf{Linear Extrapolation}, which resembles LI, but it generates data points outside the hypersphere instead of inside it. In other words, the only difference is $\alpha \sim U(1.0, a)$, for small values of $a$, so that the generated data point is close to one of the original data points used for augmentation.
    
    \item \textbf{Binary Interpolation}, which involves randomly exchanging certain features with those from another selected sample. Hence, it is similar to SP but instead of deactivating features, it replaces them with features from another sample. Hence, in the case of BI, $\alpha$ is sampled from a Bernoulli distribution $B(p)$,  $\alpha \in \lbrace 0 , 1 \rbrace^e$, $\beta = 1 - \alpha$, and $h \neq h'$.

    \item \textbf{Gaussian Scaling}, which produces scaling coefficients for each feature in the representation vector, akin to a form of perturbation noise. In this method, $h = h'$, $\alpha = 1$, and $\beta \sim N(1, \sigma)$ with small values of $\alpha$.

\end{itemize}

\section{Experimental Design}\label{sec3}

In our application of representation-level augmentation methods for vulnerability detection, we made several key decisions. Firstly, we opted to generate only vulnerable samples, given the substantial number of clean samples in our dataset. Secondly, we chose to perform all augmentations before training. This approach provides access to the entire dataset, allowing the generation of more diverse samples. In contrast to Li et al. \cite{li_exploring_2022}, we avoided in-batch augmentation during training. In-batch augmentation is commonly used in contrastive settings, where negative samples are typically generated through the augmentation process.

\subsection{Blind Augmentation} 
We applied the five augmentation methods, as introduced in Section~\ref{sec:aug_methods}, to vulnerable samples of the BigVul dataset as Blind Augmentation. This is to emphasize that the vulnerable section in each vulnerable section may
be overwritten or removed.

\subsection{Conditioned Augmentation} 
Aside from blindly applying five augmentation approaches, we also investigated the effect of fixing the vulnerable sections (i.e., conditioned augmentation) while using such augmentation methods. This involved intelligently excluding the tokens in the embedding corresponding to the specified vulnerable statement(s) within the dataset from the augmentation process. Utilizing Byte-Pair Encoding (BPE) for transforming code into tokens, we could easily identify the transformed vulnerable line in the embedding. Despite minor variations in the initial and final tokens due to the nature of BPE, we systematically searched for the sequence of tokens in the code embedding. In case of failure, we recursively excluded the first and last tokens in the line embedding until success. Subsequently, we replaced the corresponding tokens of the vulnerable line(s) to ensure the final generated vector encapsulated the vulnerability.


\subsection{Implementation Details}
To evaluate different data augmentation methods, we used the Bigvul~\cite{fan_cc_2020} dataset---as shown in Table \ref{tab:dataset}---for training and testing Linevul~\cite{fu_linevul_2022}. Specifically, we used the original training set of Bigvul and augmented the vulnerable samples of the training set that contained the vulnerable line information so that it would be a fair comparison between the naive and conditioned augmentation methods. We trained Linevul with the optimal hyperparameters mentioned in Table~\ref{tab:hyperparams}, with embedding block size and feature embedding size similar to the original \cite{fu_linevul_2022}. Feature embedding size corresponds to the vectors' size at the end of RoBERTa~\cite{RoBERTa} before being fed to the final classifier. We also used the validation set to select the best checkpoint during training for testing. To compare these methods, we employed the four common evaluation metrics \cite{EvaluationMetrics}---namely recall, precision, F1-score, and area under the curve (AUC)---in line with previous studies \cite{yang_does_2023, vuldeepecker, DeepVD, FeatureImportanceDefectClassifiers, NoiseClassifierSE, Vulscriber}.

\begin{table}[!t]
\caption{BigVul dataset}\vspace{-1em}
  \centering
  \begin{tabular}{l|c|c|c}
    \hline
    
    \textbf{Data type} & \textbf{Train} & \textbf{Validation} & \textbf{Test} \\
    \hline
    
    \textbf{Vulnerable items} & 8,783 & 1038 & 1079 \\
    \quad w/ flaw lines specified & 5,895 & - & - \\
    \quad w/o flaw lines specified & 2,888 & - & -\\
    \hline
    \textbf{Clean items} & 142,125 & 17826 & 17785\\
    \hline
    \textbf{All items} & 150,908 & 18864 & 18864\\
    \hline
    \textbf{Ratio} & 1:16.2 & 1:17.2 & 1:16.5 \\
    \hline
  \end{tabular}
  \label{tab:dataset}
\end{table}

\begin{table}[!t]
  \caption{Hyper-parameter settings for training LineVul}
\vspace{-1em}
  \centering
  \begin{tabular}{@{}c|c|c|c|c@{}}
    \hline
    \textbf{Batch} & \textbf{\#epochs} & \textbf{Learning} & \textbf{Block} & \textbf{Feature embedding}\\
    \textbf{size} & & \textbf{rate} & \textbf{size} & \textbf{size}\\
    \hline
    32 & 10 & $2 \times 10^{-5}$ & 512 & 768 \\
    \hline
  \end{tabular}
\vspace{-1em}
\label{tab:hyperparams}
\end{table}


We conducted experiments on a Linux Server equipped with two Nvidia RTX A4000 GPUs. Our configurations closely mirrored those used by LineVul \cite{fu_linevul_2022}. We employed RoBERTa \cite{RoBERTa} with pre-trained weights from CodeBERT \cite{feng_codebert_2020} attached to the default RoBERTa's classification head, and optimized the model with the Binary Cross Entropy (BCE) loss. The Byte-Pair Encoding (BPE) tokenization process was consistent across experiments for transforming code into tokens.

Aside from the two baselines of ROS and VGX, we investigate the application of the five methods outlined in Section~\ref{sec:aug_methods} in both settings (i.e., naive and conditioned):
\begin{itemize}
    
    \item \textbf{Linear Interpolation.} 
    We employed this method with elements of $\alpha$ sampled from the uniform distribution $\alpha \sim U(0.9, 1.0)$. Figure \ref{fig:blind_aug} illustrates an example of this method with the sampled $\alpha$ set to 0.95.
    
    \item \textbf{Stochastic Perturbation.} 
    In alignment with prior studies, we utilized DropOut for the implementation of this method with a $p=0.1$.

    \item \textbf{Linear Extrapolation.} 
    We employed this method with elements of $\alpha$ sampled from the uniform distribution $\alpha \sim U(1.0, 1.1)$.
    
    \item \textbf{Binary Interpolation.} 
    For this method, we opted to alter only 25\% of the cells each time, replacing them with those from another vulnerable sample, following the approach outlined by Li et al.\cite{li_exploring_2022}. 
    
    \item \textbf{Gaussian Scaling.}
    For this method, we sampled $\beta$ from a Gaussian distribution $\beta \sim N(1, 0.1)$. 

    \item \textbf{Random Oversampling (ROS).}
    We used ROS as a baseline as it can benefit DLVD and is not easy to beat~\cite{yang_does_2023,  Vulscriber}. Hence, we used ROS to sample the raw vulnerable samples randomly to get a balanced dataset.

    \item \textbf{VGX.}
    Similar to ROS, we used VGX~\cite{vgx} as another baseline and sampled the same number of vulnerable samples (i.e., 129690 items) from the 896K items provided by the original paper~\cite{vgx}.

\end{itemize}

For all of the mentioned augmentation methods in both settings, vulnerable samples were used to generate new ones; therefore, all the generated items were labeled as vulnerable. We augmented the data at a rate of 23 using all the aforementioned methods. This implies that the augmented dataset comprises the original items along with augmented data, amounting to 22$\times$ the number of vulnerable samples. This approach was adopted to achieve a balanced dataset.

\section{Results}\label{sec4}
\newcommand{\colorcell}[1]{
    \cellcolor[HTML]{#1}
}
\begin{table}[!t]
\caption{Blind augmentation results. The ones beating No augmentation are marked in blue; otherwise, marked in red.}
\vspace{-1em}
  \centering
  \begin{tabular}{@{}l@{}|c|c|c|c@{}}
    \hline
    \textbf{Augmentation strategy} & \textbf{AUC} & \textbf{Recall} & \textbf{Precision} & \textbf{F1} \\
    \hline
    No augmentation & 83.97 & 28.17 & 45.31 & 34.74 \\
    \hline
    VGX~\cite{vgx} & \colorcell{FFCCCC} 83.75 & \colorcell{FFCCCC} 26.69 & \colorcell{CCE5FF} 46.68 & \colorcell{FFCCCC} 33.96\\
    \hline
    Random Oversampling & \colorcell{99CCFF} 84.71 & \colorcell{99CCFF} 33.92 & \colorcell{FFCCCC} 44.69 & \colorcell{99CCFF} 38.5 \\
    \hline
    Linear Interpolation & \colorcell{FFCCCC} 83.53 & \colorcell{CCE5FF} 29.38 & \colorcell{99CCFF} 50.32 & \colorcell{99CCFF} 37.1 \\
    \hline
    Linear Extrapolation & \colorcell{FFCCCC}83.08 & \colorcell{99CCFF} 31.14 & \colorcell{FFCCCC} 44.80 &
    \colorcell{99CCFF} 36.74 \\
    \hline
    Stochastic Perturbation & \colorcell{FFCCCC} 83.26 & \colorcell{99CCFF} 33.09 & \colorcell{FFCCCC} 44.91 & \colorcell{99CCFF} 38.1 \\
    \hline
    Binary Interpolation & \colorcell{FFCCCC}83.69 & \colorcell{FFCCCC} 25.95 & \colorcell{99CCFF}50.18 & \colorcell{FFCCCC}34.21 \\
    \hline
    Gaussian Scaling & \colorcell{FFCCCC} 82.33 & \colorcell{CCE5FF} 28.92 & \colorcell{99CCFF}  48.15 & \colorcell{99CCFF} 36.13 \\
    \hline
  \end{tabular}
  \label{blind_aug_result}
\end{table}

\begin{table}[t]
\caption{Conditioned augmentation results. The ones beating No augmentation are marked in blue; otherwise, marked in red.}
\vspace{-1em}
  \centering
  \begin{tabular}{@{}l@{}|c|c|c|c@{}}
    \hline
    \textbf{Augmentation strategy} & \textbf{AUC} & \textbf{Recall} & \textbf{Precision} & \textbf{F1} \\
    \hline
    No augmentation & 83.97 & 28.17 & 45.31 & 34.74 \\
    \hline
    VGX~\cite{vgx} & \colorcell{FFCCCC} 83.75 & \colorcell{FFCCCC} 26.69 & \colorcell{CCE5FF} 46.68 & \colorcell{FFCCCC} 33.96\\
    \hline
    Random Oversampling & \colorcell{99CCFF} 84.71 & \colorcell{99CCFF} 33.92 & \colorcell{FFCCCC} 44.69 & \colorcell{99CCFF} 38.5 \\
    \hline
    Linear Interpolation & \colorcell{FFCCCC} 83.47 & \colorcell{99CCFF} 30.95 & \colorcell{CCE5FF} 46.39 & \colorcell{99CCFF} 37.13 \\
    \hline
    Linear Extrapolation & \colorcell{FFCCCC}83.72 & \colorcell{99CCFF} 30.77 & \colorcell{99CCFF} 48.19 & \colorcell{99CCFF} 37.56 \\
    \hline
    Stochastic Perturbation & \colorcell{FFCCCC} 83.70 & \colorcell{99CCFF}  32.81 & \colorcell{FFCCCC}44.70 & \colorcell{99CCFF} 37.84 \\
    \hline
    Binary Interpolation & \colorcell{FFCCCC}83.49 & \colorcell{99CCFF}31.33 & \colorcell{FFCCCC}44.36 & \colorcell{CCE5FF} 36.72 \\
    \hline
    Gaussian Scaling & \colorcell{FFCCCC}83.43 & \colorcell{99CCFF} 30.58 & \colorcell{CCE5FF}47.08 & \colorcell{99CCFF}37.07 \\
    \hline
  \end{tabular}
\vspace{-1em}
  \label{conditioned_aug_result}
\end{table}

\textbf{\textit{Finding 1:} Stochastic Perturbation performs better than other token-based representation-level methods in both settings.}
Tables~\ref{blind_aug_result} and \ref{conditioned_aug_result} show the results of the five representation-level augmentation methods when applied blindly (without conditioned augmentation) and conditioned, respectively. We observe that the majority of such augmentation methods lead to performance improvements as they are applied blindly. Binary Interpolation is the only method that degrades the overall F1-score, and Stochastic Perturbation generates the highest gain (9.67\%). We observe a similar trend in conditioned augmentation variants, where all methods achieved a performance gain compared with no augmentation, and Stochastic Perturbation performs the best among the conditioned representation-level augmentation approaches as well and achieves an improvement (8.92\%) over No Augmentation in terms of F1-score.

\textbf{\textit{Finding 2:} None of the representation augmentation methods can beat the ROS.} Interestingly, when compared with the representation augmentation methods with both blind and conditioned settings, ROS outperforms all of them, with an improvement of 10.82\% in terms of F1-score, compared with No Augmentation. 

\textbf{\textit{Finding 3:} All representation-level augmentation methods beat VGX, a SOTA vulnerability generation method.} Intriguingly, VGX (a SOTA vulnerability generation method that outperforms Vulgen) fails to increase performance when used for balancing a real-world dataset like BigVul, which is more imbalanced and bigger than other datasets such as Devign~\cite{Devign} and Reveal~\cite{Reveal}.





\section{Discussion}\label{sec5}

\subsection{Why Does Stochastic Perturbation's Performance Degrade When Conditioned?}
As we delve into the mechanism of \textbf{Stochastic Perturbation}, we realize that using this method to augment data is equivalent to using a DropOut layer in the network, which gets omitted one out of 23 times (the original data) it sees a vulnerable sample. However, in the conditioned setting, since the vulnerable tokens will be reinstated after the DropOut, they will behave differently from the original DropOut's regularization effect, and the expected value of the output will no longer be equal to that of the input data. This is while other methods do not have such an effect in the naive setting, and keeping the vulnerable section proves to be useful for them.

\subsection{Why None of the Methods Can Beat ROS?}
It is known that noise in the data can act as a regularizer for machine learning models~\cite{NoisyDataTikhonovRegularization}. Hence, if the data is too noisy, then the model's learning process will be affected and limited. In the case of studied methods, all of these methods generate samples that contain numbers that do not necessarily mean anything, and so the generated sample is very noisy such that simply repeating the minority class will have a better effect on the learning of the model.

\subsection{Why Does VGX Perform Worse Than All Representation-Level Augmentation Methods?}
Although VGX is a vulnerability generation method and is expected to generate samples close to real code, this is not the case. The VGX's output requires manual checking to remove invalid code samples. Also, it is worth noting that VGX only targets single statement vulnerabilities which do not present the majority of vulnerabilities that span multiple lines. Hence, VGX is only useful when used in moderation and there are experts available for manual checking. Hence, if we use VGX without manual checking, too much noise will be added while changing the distribution of vulnerabilities. This guides the model to focus more on single-statement vulnerabilities, and too much noise prevents proper learning of the models.

\subsection{Implications of Our Findings}

We recommend that future researchers consider using \textbf{Random Oversampling} instead of token-based representation-level vulnerability augmentation when aiming for a balanced dataset. However, one might prefer to keep the original ratio and augment the minority class while adding new items of the majority class that are widely available. In such a case, \textbf{Stochastic Perturbation} seems the best augmenting strategy to create new items. This is because adding new clean items while repeating the same vulnerable samples will only add the unwanted noise of the new clean items, without adding any extra useful information about the vulnerabilities. Also, based on the results of VulScriber~\cite{Vulscriber} and ours, we conclude that VGX~\cite{vgx} is not a practical method for dealing with data shortage when we aim for either balancing a dataset or keeping the original ratio. 

Lastly, our examination of conditioned methods reveals that while they offer potential, they still introduce a notable amount of noise. Given these observations, we advocate for future investigations to delve deeper into more intrusive augmentation techniques for DLVD. This recommendation contrasts with prior studies that found success with similar methodologies for different software engineering tasks.

\subsection{Threats to Validity and Limitations}
\textbf{Internal Validity:} An internal threat relates to the conditioned augmentation methods. We hypothesized that segments containing vulnerable tokens are sensitive to alterations during the blind augmentation. The random variation introduced to the corresponding tokens might negate their vulnerability representation. This assumption is derived from observations suggesting that modifying a single syntax symbol in a code snippet could either eliminate the vulnerability or introduce new forms of errors. However, validating this assumption experimentally proves challenging. This limitation arises from the inherent blind nature of representation-level augmentation, as newly generated tokens may not map back to meaningful source code, let alone compilable code.

\textbf{External Validity:} An external threat is associated with the dataset utilized for training and testing our model, as our experiments exclusively involved C++ code snippets. This choice may present challenges to the generalization capabilities of our augmentation methods. 

Additionally, a threat concerns the generalization capacity of our method, which is influenced by the choice of the model. Since our study focuses on augmentation, and many algorithms are tested concurrently, our analysis does not cover a wide range of Vulnerability Detection models. We exclusively employed LineVul (a recent DLVD model) due to its high performance, partly influenced by time constraints and resource availability. In the future, we plan to include other models to validate the effect of our augmentation methods.


\section{Conclusions}\label{sec6}
In this paper, we conducted the first comprehensive study on the impact of representation-level augmentation in the context of Deep Learning-based Vulnerability Detection. Our study introduced a vulnerability detection-specific approach to augmenting data, conditioning the augmentation based on the vulnerable segment in the embedding. Our findings offer valuable insights and recommendations for practitioners in this field. Notably, Stochastic Perturbation emerged as the top-performing method. We also showed that conditioning the augmentation is beneficial for most representation-level augmentation methods.

As for future work, we plan to investigate the effect of graph-based augmentation methods (e.g., Manifold Mixup) that, aside from the text representation, also take the graph representation of the code (e.g., AST or CFG) into account.  

\section*{Code Availability}\label{sec7}
Our code is available at \href{https://zenodo.org/records/13916933}{https://doi.org/10.5281/zenodo.13916933} for replication and further improvements.

\section*{Acknowledgments}
This work is partially supported by NSERC (Canada) and the University of Manitoba.


%


\bibliographystyle{ACM-Reference-Format}
\bibliography{ref}


\end{document}